\documentclass[sigconf]{acmart}
\pdfoutput=1
\usepackage{listings}
\usepackage{lstnix}
\usepackage{lstjson}
\usepackage{xcolor}
\usepackage{adjustbox}
\usepackage{todonotes}
\usepackage{caption}
\usepackage{float}
\usepackage{color}
\usepackage{xcolor}
\usepackage{lipsum}
\usepackage{subcaption}
\usepackage[htt]{hyphenat}
\usepackage{algorithm}
\usepackage[noend]{algpseudocode}
\usepackage{siunitx}

\newcommand{\nixpkgs}{Nixpkgs}

\acmConference[ICSE '24]{International Conference on Software Engineering}{April 14-20, 2024}{Lisbon, Portugal}

\acmPrice{}
\copyrightyear{2024}
\acmYear{2024}
\setcopyright{licensedothergov}\acmConference[ICSE-NIER'24]{New Ideas and Emerging Results }{April 14--20, 2024}{Lisbon, Portugal}
\acmBooktitle{New Ideas and Emerging Results (ICSE-NIER'24), April 14--20, 2024, Lisbon, Portugal}
\acmDOI{10.1145/3639476.3639767}
\acmISBN{979-8-4007-0500-7/24/04}

\begin{document}

\title{Reproducibility of Build Environments through Space and Time}

\author{Julien Malka}
\email{julien.malka@ens.fr}
\orcid{0009-0008-9845-6300}
\affiliation{\institution{LTCI, Télécom Paris, \\Institut Polytechnique de Paris}
 \city{Palaiseau}
 \country{France}
}
\author{Stefano Zacchiroli}
\email{stefano.zacchiroli@telecom-paris.fr}
\orcid{0000-0002-4576-136X}
\affiliation{\institution{LTCI, Télécom Paris, \\Institut Polytechnique de Paris}
 \city{Palaiseau}
 \country{France}
}
\author{Théo Zimmermann}
\email{theo.zimmermann@telecom-paris.fr}
\orcid{0000-0002-3580-8806}
\affiliation{\institution{LTCI, Télécom Paris, \\Institut Polytechnique de Paris}
 \city{Palaiseau}
 \country{France}
}

\begin{abstract}

Modern software engineering builds up on the composability of software components, that rely on more and more direct and transitive dependencies to build their functionalities. This principle of reusability however makes it harder to reproduce projects' build environments, even though reproducibility of build environments is essential for collaboration, maintenance and component lifetime. In this work, we argue that functional package managers provide the tooling to make build environments reproducible in space and time, and we produce a preliminary evaluation to justify this claim. Using historical data, we show that we are able to reproduce build environments of about 7 million Nix packages, and to rebuild 99.94\% of the 14 thousand packages from a 6-year-old \nixpkgs\ revision.

\end{abstract}

\maketitle

\section{Introduction}

Modern software engineering is built on the principles of \textbf{composability} and \textbf{reusability}: instead of re-implementing existing functionalities, software typically depends on libraries providing them. This approach increases productivity and helps build more robust software: each component focuses on a little piece of functionality, and tries to do it well. However, as a consequence, software projects accumulate (direct and transitive) dependencies, resulting in an increasingly complex software supply chain~\cite{kikas_structure_2017}. \textbf{Build environments}---that is, all the components and variables, up to their specific versions, necessary to build the software---become more complex and difficult to reproduce in space and time, a situation often referred to as \textit{dependency hell}~\cite{dick_dll_2018}. 

We say that a build environment is reproducible \textit{in space} when it is possible to obtain an environment containing the exact same components and variables on another machine. We say that it is reproducible \textit{in time}, when this property does not decay over time, which means one can obtain a build environment that remains identical to the one that was used when building a given software version in the past. Reproducibility of the build environment is a desirable property for a software project: in space, it allows for smoother collaboration between developers that can then work on the project in identical conditions, thus alleviating barriers to contribution. It also facilitates bug reproduction by projects maintainers. Reproducibility of the build environment in time allows rebuilding past software versions, which is very useful to better understand how software used to work or to bisect bugs~\cite{courtiel_theoretical_2022, lwn_git_bisect_run}. It also helps combat the risk of software dying because progressive independent evolution of its dependencies makes it practically impossible to find a combination of dependency versions that allows rebuilding it.

A number of package managers have taken steps towards improving the reproducibility of their build environments by introducing \textit{lockfiles} that contain a reference to the exact versions of dependencies used by the project (e.g., npm with \texttt{package-lock.json}~\cite{goswami_2020_npm_pkg_repro}). Although they do help, these methods are insufficient for projects that have dependencies beyond a single ecosystem. Another popular approach is to provide a container image, allowing users to reproduce the same build environment by downloading the image and running the container. However, making the process of generating these images reproducible itself requires care~\cite{nust_ten_2020}. 

We believe that making build environments easily reproducible could have decisive impact on the practice of software engineering by facilitating collaborative development, helping maintenance and reinforcing software composability by combatting dependency aversion. We also aim to demonstrate that functional package managers, like Nix~\cite{dolstra_purely_2006} or Guix~\cite{courtes_2013_guix}, are the right tools for this task.
While the Nix and Guix communities claim that they provide the necessary tooling to achieve reproducibility of build environments~\cite{pjotr_nix_2008}, no empirical evidence has been available thus far to support these claims. In this work, we empirically evaluate whether space and time reproducibility of build environments is achievable using the Nix package manager (\textbf{RQ1}) and if that allows rebuilding past software versions (\textbf{RQ2}). While making build environments reproducible may help perform \textit{bit-by-bit} reproducible builds~\cite{lamb_2022_reproducible_builds}, the study of Nix abilities for that purpose is left for future work. 
 
Our results show that we can achieve 99.99\% reproducibility of build environments over \num{7010516} packages coming from 200 historical revisions of \nixpkgs, the Nix package set. Additionally, we were able to rebuild 99.94\% of the packages from a 6-year-old \nixpkgs\ revision, demonstrating that reproducibility of build environments is actually useful for software rebuildability.

\section{Context \& definitions}

\paragraph{Functional Package Manager (FPM)} Nix and Guix are implementations of a package deployment model, first introduced by Dolstra~\cite{dolstra_purely_2006}, that is conceptually very different from most other package managers.
In FPMs, packages are distributed as \textbf{pure functions} of their build- and run-time dependencies. In Nix, for example,
packages are specified as expressions in the Nix language. Figure~\ref{fig:nix_expression} shows a Nix expression for the \texttt{nano} text editor. This is a function whose inputs \boxed{1} are: \texttt{stdenv} (minimal build environment), \texttt{fetchurl} (function to download the program sources), and \texttt{curses} (a build-time dependency).
The output of the \texttt{stdenv.mkDerivation} function~\boxed{2} is a \textbf{derivation}, the intermediate representation Nix will use to build the package. The derivation is constructed by passing several arguments to the \texttt{mkDerivation} function: the \texttt{src} parameter \boxed{3} contains the source of the software up to its specific version (Nix checks that the correct version has been downloaded by comparing the hash of the content with the specified hash). The build-time dependencies are specified in the \texttt{buildInputs} list \boxed{4}. The \texttt{ncurses} object is also a Nix derivation, that Nix will build \textit{before} \texttt{nano}. A Nix derivation is simply a \textit{build recipe}, that Nix will use to create a \textit{build environment} with the \texttt{source} and \texttt{buildInputs} available in it. It will then run a bash script called the \textit{builder} to produce the final artifact. It is possible to provide a custom builder, but here the default autoconf builder is used, and some flags are passed in \boxed{5}. 

\begin{figure}
\begin{adjustbox}{max width=\linewidth}
\begin{lstlisting}[language=Nix, escapechar=|]
{ stdenv, fetchurl, ncurses }: | \boxed{1} |
stdenv.mkDerivation rec { |\boxed{2} |
  pname = "nano";
  version = "7.2";
  src = fetchurl { |\boxed{3} |
    url = "mirror://gnu/nano/${pname}-${version}.tar.xz";
    sha256 = "hvNEJ2i9KHPOxpP4PN+AtLRErTzBR2C3Q2FHT8h6RSY=";
  };
  buildInputs = [ ncurses ]; |\boxed{4} |
  configureFlags = [ "--sysconfdir=/etc" ]; |\boxed{5} |
}
\end{lstlisting}
\end{adjustbox}
\caption{Nix expression of nano, modified for readability.}
\label{fig:nix_expression}
\end{figure}
\begin{figure}
\begin{adjustbox}{max width=\linewidth}
\begin{lstlisting}[language=Json, escapechar=|]
{
 "args": [ "-e", "/nix/store/6xg25947...-default-builder.sh" ], |\boxed{6} |
 "builder": "/nix/store/rhvbjmcf...-bash-5.2-p15/bin/bash",
 "env": {
  "buildInputs": "/nix/store/9jmgsy8b...-ncurses-6.4-dev",
  "cmakeFlags": "",
  "configureFlags": "--sysconfdir=/etc",
  "name": "nano-7.2"
 },
 "inputDrvs": { |\boxed{7} |
  "/nix/store/3qsdhv4v...-stdenv-linux.drv": [ "out" ],
  "/nix/store/asq3sjwr...-nano-7.2.tar.xz.drv": [ "out" ],
  "/nix/store/had1mg70...-bash-5.2-p15.drv": [ "out" ],
  "/nix/store/q56mxpcf...-ncurses-6.4.drv": [ "dev" ]
 },
 "inputSrcs": [ "/nix/store/6xg25947...-default-builder.sh" ],
 "outputs": {
  "out": { "path": "/nix/store/385vk5j4...-nano-7.2" } |\boxed{8} |
 },
}
\end{lstlisting}
\end{adjustbox}
\caption{Derivation of nano, modified for readability.}
\label{fig:drv}
\end{figure}

The process by which Nix creates a derivation from a Nix expression is called \textbf{evaluation}. The resulting derivation can then be used to build a program. The outputs of both the evaluation phase and the build phase are stored in the special \texttt{/nix/store} directory of the build host. 
The name of derivation files contain a cryptographic hash that is computed based on its contents (including the precise version of the program sources, but also of all its dependencies). This use of cryptographic hashes for paths in the Nix store allows it to contain many versions of the same package.

Figure \ref{fig:drv} shows an extract of the content of the derivation obtained by evaluating the previous Nix expression.
It contains all the information derived from the Nix file. We can find the path to the builder \boxed{6}, the list of the derivations on which \texttt{nano} depends \boxed{7}, including one for the program sources, and the \texttt{/nix/store} path where the build process will create its outputs \boxed{8}, which also contains a cryptographic hash that depends on the exact dependency versions and build parameters. For details about how the outputs hashes are computed see Dolstra~\cite{dolstra_purely_2006}.

Derivations can be used in two ways: they can be instantiated, which means running the builder in the specified (hermetic and sandboxed) build environment to produce a build output in the Nix store, or they can be used to spawn a build environment where developers can then manually run their builds. One of the main claims associated with FPMs is that this build environment will be highly reproducible and that it will therefore allow performing the same build reliably (from one machine to another, and over time).

\paragraph{\nixpkgs, the Nix package collection}

The users of Nix have collaboratively created a collection of Nix expressions to build various pieces of software. This collection forms the basis of the NixOS Linux distribution~\cite{dolstra_2010_nix}, but it is available beyond NixOS, to any Nix user. With over \num{80000} packages, \nixpkgs\ is, at the time of writing, the largest Linux distribution in number of packages~\cite{repology}, as it repackages many pieces of software coming from application-specific (e.g., Emacs, VS Code) and programming-language specific (e.g., Haskell, Python) ecosystems. \nixpkgs\ provides several ``channels'', including a rolling-release called \texttt{nixpkgs-unstable}.

\paragraph{Store substitution mechanism}

While \nixpkgs\ is a source-based software distribution, Nix allows to substitute build outputs resulting from the build process specified in derivations with pre-built outputs provided by a binary cache. A binary cache is a large dictionary linking output paths to compressed build outputs. When Nix is configured to use a binary cache (a.k.a., a substituter), Nix will pause at the end of the evaluation phase and query the binary cache for the output paths of the obtained derivations. When such output paths are available in the binary cache, Nix will then download and unpack them in the Nix store, instead of running the derivation build process. It will then proceed to build the remaining derivations, for which no cached artifact was available.

\paragraph{Hydra, the Nix continuous integration platform}

The \nixpkgs\ distribution comes with an official binary cache, \texttt{cache.nixos.org}, which is populated by Hydra, a continuous integration platform. At regular intervals of time, Hydra evaluates the current revision of the \nixpkgs\ git repository. This evaluation results in a list of derivations (otherwise called \textit{jobs}) to build. Hydra then builds all of these jobs, unless they produce an output path already in cache. If the build is successful for a pre-defined list of important jobs, Hydra then pushes the build outputs to the official binary cache and updates the \texttt{nixpkgs-unstable} channel.

\section{Methodology}

In this section we describe the methodology followed to answer the stated research questions.

\subsection{RQ1: Reproducibility of build environments}

To evaluate whether Nix build environments are reproducible in space and time, we focus on the Nix evaluation step. As explained before, evaluation is the process by which Nix transforms an expression written by humans into a machine-readable derivation, that defines exactly how to create a build environment, the versions of dependencies to put in scope, and how to set environment variables. Nix users can then use the derivation to spawn a build environment, or let Nix perform the build of a known software version.

Using historical data coming from Hydra, we are interested in assessing whether we can reproduce:
the exact same list of jobs resulting from the evaluation of a given \nixpkgs\ revision (\textbf{RQ1.1});
and identical derivations for each of these jobs (\textbf{RQ1.2}).

Comparing derivations that we obtain locally with the historical ones from Hydra is actually difficult, because Hydra does not make available the derivations it built (it only pushes build outputs to the binary cache). Sometimes, we can get the path of the derivation, and that would be sufficient to check if the locally built and Hydra derivations are identical or differ, since the path contains a cryptographic hash computed from the content of the derivation. But even this path is not always available, because Hydra does not keep its trace when it did not rebuild a job (whose output was already in cache).
Besides, it can happen that two derivations with a different hash share the same output path. Indeed, derivation hashes incorporate more information on the build process than output paths (including, e.g., what \texttt{curl} version to use to download the program sources). This is because the computation of output paths is designed to create identical paths when differing derivations are guaranteed to produce the same result. Given the way output paths are computed, identical output paths should still ensure identical build environments. Therefore, we adjust \textbf{RQ1.2} slightly into: assessing whether we can reproduce identical output paths for all the jobs (as we can always retrieve historical Hydra output paths).

We perform our experiment on a sample of the \nixpkgs\ revisions. We start from a dataset (available at \href{https://channels.nix.gsc.io}{\color{blue}channels.nix.gsc.io}), which contains more than 2200 revisions belonging to the \texttt{nixpkgs-unstable} channel spanning from 2017 to 2023, with in average 23 hours of separation between each of them. To keep our study computationally reasonable, we extract 200 revisions from this dataset, keeping at the same time the maximum time spread possible and a regular spacing between the selected revisions.
The result of the sampling operation is a set of 200 \nixpkgs\ revisions that:
\begin{itemize}
    \item have been promoted to the \texttt{nixpkgs-unstable} channel;
    \item span from 2017 to 2023;
    \item have a regular spacing of 10 days and 20 hours on average.
\end{itemize}

We evaluate the selected revisions using a variant of the Nix evaluator called \texttt{nix-eval-jobs}~\cite{nix-eval-jobs}, a piece of software derived from Hydra's component in charge of the evaluation phase. It allows for faster evaluation of complete \nixpkgs\ revisions than the built-in Nix evaluator as it can evaluate several Nix derivations in parallel, and, contrary to the built-in Nix evaluator, it will not fail because some jobs fail to evaluate. This is important when building entire \nixpkgs\ revisions as there are always a few jobs that fail to evaluate, sometimes on purpose (e.g., because they have security issues).

We scrape Hydra's website to obtain historical results to compare to. For each sampled revision, we scrape from the associated Hydra webpage the list of jobs that succeeded to evaluate, and for each of them, we scrape the job webpage to retrieve its output path(s).

\subsection{RQ2: Rebuilding past software versions}

We are interested in understanding if achieving build environment reproducibility is sufficient to confidently rebuild past software versions. To answer this, we build all the jobs from our most ancient \nixpkgs\ revision (from 2017, which is the oldest \texttt{nixpkgs-unstable} revision available in our dataset). It contains \num{14753} jobs for the \texttt{x86\_64-linux} architecture, out of them \num{14461} built successfully at the time. We compare the build status stored in Hydra to our local build success or failure. In case a build fails in our experiment while it had succeeded on Hydra in 2017, we look at the build log to try to understand the cause of the failure.

Note that, as this step is much more computationally intensive than the previous one (we are actually building an entire distribution of open source software), our sample size is much smaller. Instead of evaluating the reproducibility over time on a large sample of \nixpkgs\ revision, we focus on the oldest revision as we expect that, in principle, the farther away we go in the past, the more difficult it becomes to rebuild software. Therefore, we expect that if we obtain good results on this old revision, they should reasonably extend to more recent revisions. We recognize that there are threats to this claim of external validity, e.g., if changes in software practice or in the scope of \nixpkgs\ trigger an increase in flaky builds.

Besides, for now, we do not check for \textit{bit-by-bit} reproducible builds~\cite{lamb_2022_reproducible_builds}. Extending our evaluation to build more revisions and checking for reproducible builds are both part of our future plans.

We use the Nix cache extensively in this step. This allows building many jobs in parallel, by focusing on rebuilding the selected job and not its dependencies. While this helps answer the question ``Is build environment reproducibility sufficient for rebuilding past software versions?'', it does not answer the question ``Can we rebuild an entire past revision of \nixpkgs\ without relying on cache?''. The latter would be a much more difficult achievement because failures would induce cascade effects on all their dependents, but also because of the risk of program sources becoming unavailable. Currently, these sources are still provided through the Nix cache.

\section{Results}

In this section, we describe the results obtained by performing our experiments.

\subsection{RQ1: Reproducibility of build environments}

During the first phase of our experiments to assess the reproducibility of the evaluation of \nixpkgs\ revisions, we discovered two bugs that resulted in obtaining different lists of jobs. One was a bug of \texttt{nix-eval-jobs}, which did not follow the current convention in \nixpkgs\ and Hydra to determine when to include a job or not in the result of the evaluation. We fixed this bug (see~\cite{malka_fix_nodate}) and we used the updated version to perform our experiments. The other was a bug of Hydra, which skipped jobs which contained a dot in their name (which is quite rare in \nixpkgs). We also fixed this bug (see~\cite{malka_hydra-eval-jobs_nodate}), but we had to take this bug into account when performing our comparisons with historical Hydra data.

Up to the discrepancy coming from this second bug, our results show that we can obtain 100\% identical lists of jobs from our sample of historic \nixpkgs\ revisions.
When comparing jobs' output paths, we obtain 99.99\% identical output paths. For each revision, there were at most 4 expressions causing jobs with differing output paths and they all used Nix features that cause impure evaluation (\texttt{builtins.nixVersion} or \texttt{lib.inNixShell} for example).
These results show that Nix is able to achieve perfect build environment reproducibility, given an unchanged Nix expression (avoiding impure features), and despite several differences in the conditions of the evaluation: different hardware, Linux distribution (NixOS for the Hydra infrastructure, Ubuntu for our experiments), different Nix versions (we used Nix version 2.6 in our experiments, while the Nix version used in Hydra has evolved over the years), and points in time (between 0- and 6-year differences).

\subsection{RQ2: Rebuilding past software versions}

We were able to successfully rebuild \num{14452} out of the \num{14461} jobs that Hydra had successfully built from our selected revision (\texttt{5328102}), giving us a success rate of 99.94\%. We performed a preliminary assessment of the reasons for the failures of the 9 remaining jobs, and confirming or adjusting this assessment is part of our future plans. At this point, we have classified the jobs into 3 classes of reasons that might cause the failure.

\paragraph{Current build sandbox leakage} 3 jobs failed because the build script rejected the kernel version or the OS used, pieces of information that should not be available inside the sandbox.
\paragraph{Flaky tests} 1 job failed because of a single failed test, which makes us suspect a flaky test (a test that fails inconsistently)~\cite{luo_empirical_2014, parry_survey_2021}.
 \paragraph{Past build sandbox leakage} For several other jobs, including 1 whose tests fail because of an expired certificate, and 2 whose failures look related to the shell, we suspect that the failure could come from a stricter build sandbox in newer Nix versions, preventing the build to have access to unspecified dependency or data that would have previously been available from the environment.

\section{Related work}

``Moving parts'' in build environments have been recognized as problematic for a long time, because they lead to non-reproducibility issues.
Continuous Integration (CI)~\cite{duvall_continuous_2007, meyer_continuous_2014}, a key DevOps practice, expects build environment stability.
CI build failures have been studied empirically and at a large scale~\cite{seo_programmers_2014, rausch_empirical_2017, zhang_large-scale_2019}.
It is well-established that, independently of the programming language, the most common cause of CI build failures are (bloated) dependency issues~\cite{soto-valero_comprehensive_2021}.
More generally, the so-called ``dependency hell''~\cite{dick_dll_2018} is a major factor in the non-reproducibility of development environments. For example, Mukherjee et al.~\cite{mukherjee_fixing_2021} investigated how this is the case in the Python ecosystem; Zampetti et al.~\cite{zampetti_empirical_2020} confirm this in a broader study of ``CI smells''. 
Abate et al.~\cite{abate_2015_msr_installability_issues, goswami_2020_npm_pkg_repro} show how failed dependency resolution is a recurrent cause of package non-installability across different package ecosystems.
Flaky tests~\cite{luo_empirical_2014, parry_survey_2021} are equally annoying for developers and can also be caused by the unexpected displacements of build environment parts, including dependencies.
Dependency pinning, as supported by package manager ``lockfiles'', is a partial solution to the problem~\cite{goswami_2020_npm_pkg_repro}, which does not address reproducibility causes other than moving dependencies.

In 2020, the ReScienceC journal ran the Ten Years Reproducibility Challenge~\cite{rescience_repro_challenge, perkel_challenge_2020}, defying scientists into rerunning software associated to their own papers published at least 10 years prior.
Participants generally succeeded, but reported about significant difficulties in doing so.
Supporting reproducible science via software tooling is currently a hot topic, with researchers looking into how to leverage Docker for that~\cite{boettiger_2015_docker_repro_research, cito_2016_docker_repro_emse}, but also functional package managers~\cite{strozzi_2019_repro_genomics_guix, courtes_2015_repro_hpc_guix}.
One of the promises of functional package management~\cite{dolstra_purely_2006, courtes_2013_guix} is indeed to fully describe build environments, removing dependency issues from the equation of build and test failures.
This is the \emph{theory}, at least, but one that to the best of our knowledge had never been empirically validated before.

Build environment reproducibility is also important for software preservation~\cite{shustek_what_2006}.
Previous works have observed~\cite{matthews_towards_2009, satyanarayanan_saving_2018} that build systems, recipes, and tools, need to be preserved as much as source code~\cite{di_cosmo_software_2017} and binary executables.
But preservation is less useful if, after having captured all of that, the build process cannot be replicated to obtain the intended result.

\section{Future plans}

This study brings preliminary evidence that Nix allows specifying build environments that are reproducible both in space and time, and that most often, this enables rebuilding past software versions. We intend to complete this initial work by exploring more largely our dataset in order to understand the limits of this property: by rebuilding more \nixpkgs\ revisions, we will observe the evolution over time of package rebuildability. We have used Nix 2.6 to perform our experiments and relied on the fact that Hydra used diverse Nix versions over time to claim that Nix achieves build environment reproducibility with different Nix versions, but it would strengthen our results to make the local version of Nix vary as well. Since we collected logs of our local rebuilds and that Hydra's build logs are also available, we can analyze them to understand why some packages fail to rebuild, but also to measure how often the build logs are fully identical. We have argued that Nix reproducible environments enable rebuilding a component long after it was defined, but we have only tested running the build step, with all dependencies pulled from the Nix binary-cache. More failures are to be expected if we do not rely on a binary-cache, in particular when the program sources become unavailable. Because we are interested in the question of software preservation, we plan to evaluate the amount of packages that we are not able to build because of sources unavailability and the proportion we can salvage using sources archives like Software Heritage~\cite{di_cosmo_software_2017}. Finally, we are interested in the abilities of functional package managers for reproducible builds, for the impacts it can have on the security of the software supply chain~\cite{lamb_2022_reproducible_builds}. We anticipate that the reproducibility of build environments enabled by Nix can help achieve reproducible builds, but the effect of other factors like compiler behaviors and quality of packaging are also to be considered.

\paragraph{Data availability} 
A full replication package for the experiments described in this paper is available from Zenodo~\cite{malka_replication_2024} and archived on Software Heritage with SWHID \\
\href{https://archive.softwareheritage.org/browse/revision/f513eee162ea28ab3066eb1c0aac57b80f16cc5c/}{\texttt{swh:1:rev:f513eee162ea28ab3066eb1c0aac57b80f16cc5c}}.
\clearpage


\end{document}